\documentclass[aps,preprint,showpacs,preprintnumbers,amsmath,amssymb]{revtex4}
\usepackage{amsmath,mathrsfs,amsbsy,color,graphicx,bm,amsthm,amsfonts}
\usepackage{units}
\usepackage{bbm}
\usepackage{times}
\usepackage{dcolumn}
\usepackage{mathrsfs}% Align table columns on decimal point
\usepackage{amsmath,amssymb,epsfig}

\begin{document}

\title{Quantum correlation between a qubit and a relativistic boson in an expanding spacetime}
\author{Shu-Min Wu$^1$, Hao-Sheng Zeng$^2$\footnote{Email: hszeng@hunnu.edu.cn}, Tonghua Liu$^{3}$}
\affiliation{$^1$ Department of Physics, Liaoning Normal University, Dalian 116029, China\\
$^2$ Department of Physics, Hunan Normal University, Changsha 410081, China\\
$^3$ Department of Physics, Yangtze University, Jingzhou 434023, China}

% \baselineskip=0.65 cm

%\vspace*{0.2cm}
\begin{abstract}
We use the quantum correlation of both logarithmic negativity and mutual information between a qubit and a relativistic boson to analyze the dynamics of universe expansion. These dynamical quantum correlations can encode the information about underlying spacetime structure, which suggests a promising application in observational cosmology. We find that the dynamics of both logarithmic negativity and mutual information between the qubit and the boson are very similar. They decrease monotonically with the growth of the expansion volume and the expansion rate. Smaller momentum and medium-sized mass of boson are more favourable for extracting the information about history of universe expansion. The quantum correlation between the qubit and the antiboson however has very different behavior: The logarithmic negativity is always zero and the mutual information can be generated through the expansion of universe. Smaller momentum and medium-sized mass of antiboson are beneficial for the production of mutual information. Finally, the trigger phenomenon and conservation for mutual information are witnessed.

\end{abstract}

\vspace*{0.5cm}
\maketitle
\section{Introduction}
Quantum entanglement plays an important role in quantum information theory.
It is a resource for quantum information tasks, such as quantum communication,
teleportation and various computational tasks, and has applications in quantum control
and quantum simulations \cite{L1,L2,L3,L4}. Though the study about quantum entanglement is originated initially from the nonrelativistic realm, extending it to the relativistic domain is ultimately necessary because the world can never escape the influence of gravity. In fact, many settings, such as photons in the realization of quantum information \cite{L5,L6,L7,L8,L9,L10,wsm1,L11,L12,L13,L14,L15,L16,L17,LL15,wsm2,wsm3}, involve relativistic effects.
Moreover, quantum entanglement plays a prominent
role in the information loss problem and in black hole thermodynamics \cite{L18,L19,L20,L21}.

In fact, the notion of relativistic quantum information includes two branches: the special relativity effect \cite{Moradpour2015} and the general relativity (acceleration and gravity) effect. In the latter case, people mainly concern two aspects:
(i) Study the influence of gravity or acceleration on quantum resources or quantum settings, such as quantum entanglement, quantum coherence, quantum key distribution and quantum communication \cite{L5,L6,L7,L8,L9,L10,wsm1,L11,L12,L13,L14,L15,L16,L17,LL15,wsm2,wsm3}; (ii) Probe the property of spacetime via quantum effect \cite{L22,L23,L24}. Quantum  entanglement subject to gravitational effect shows that conceptually important qualitative differences from the nonrelativistic perspective arise.
For example, quantum entanglement depends on the motional state of the observer and degrades from the
perspective of accelerated observers \cite{L5,L6,L7,L8,L9,L10}.  These results imply that quantum resources
in relativistic settings might not be an invariant concept.

In this paper, we consider an entangled state between a qubit and a relativistic boson. The latter is transformed, via the expansion of a Robertson-Walker spacetime, to a superposition of different number states of boson and antiboson.
We then study, in the final state, the pairwise quantum correlation (i.e.,logarithmic negativity and mutual information) between the qubit and the boson, or between the qubit and the antiboson. The dynamics, redistribution and conservation of the underlying quantum correlations in the expansion of universe are our research focuses. Such researches not only deepen our understanding
about quantum correlations but also provide the prospect of employing quantum correlations as a tool to study the property of curved spacetime, because the dynamical quantum correlations contain information about the history of the expanding spacetime, affording the possibility of deducing cosmological parameters of the underlying spacetime.
This novel way of obtaining information
about cosmological parameters could both theoretically and experimentally  provide new insight
into the early universe \cite{LL24,LLL24}.
Actually, it has been shown that quantum correlations play an important role in the thermodynamic properties of Robertson-Walker type spacetime, and can be used to distinguish different spacetimes, and probe
spacetime fluctuations \cite{L25,L26,L27}.

The paper is organized as follows. In Sec. II, we
present curtly the quantization of scalar fields in expanding spacetime. Sec. III is the main part of the paper, where the dynamics, redistribution and conservation of quantum correlations in the expanding spacetime are investigated.  The last section is devoted to the brief conclusion.
%------------------------------------------------------------------------------------------------------------------------------------------------------------------------------------------------%
\section{Quantization of scalar fields in expanding spacetime \label{GSCDGE}}
%--------------------------------------------------------------------------------
Let us consider a two-dimensional Robertson-Walker spacetime with the
line element described by the metric \cite{L28,L29,L30}
\begin{equation}\label{F1}
ds^2=dt^2-[a(t)]^2 dx^2\,,
\end{equation}
where $a(t)$ is the scale factor. Introducing the conformal time $\eta$ that is related to the cosmological time $t$ by
\begin{equation}\label{F2}
\eta=\int_0^t \frac{\text{d}\tau}{a(\tau)}\,,
\end{equation}
then the metric has the form
\begin{equation}\label{MconFLRW}
ds^2=[a(\eta)]^2(d\eta^2- dx^2)\,.
\end{equation}
Following the standard notation, the conformal scale factor $C(\eta)=[a(\eta)]^2$ has specific form
\begin{equation}\label{factorbos}
C(\eta)=1+\epsilon(1+\tanh(\rho\eta))\,,
\end{equation}
where $\epsilon$ is positive real parameter controlling the total expansion volume, and $\rho$ is positive real parameter controlling the expansion rate.
The conformal scale factor $C(\eta)$ conveniently shows that
the spacetime undergoes a period of smooth expansion for the universe, and becomes flat in the distant past $\eta\rightarrow -\infty$ [$C(\eta\rightarrow -\infty)=1$] and in the far future $\eta\rightarrow \infty$ [$C(\eta\rightarrow \infty)=1+2\epsilon$]. For convenience, the distant past and far future are referred to as the asymptotic ``in"-region and ``out"-region, respectively.

In the background of the expanding spacetime,  a bosonic field obeys the Klein-Gordon equation
\begin{equation}\label{F3}
[\partial_\eta^2+k^2+C(\eta)m^2]\Phi=0.
\end{equation}
Solving this equation, we obtain two sets of complete
modes in the asymptotic past and the asymptotic future
\begin{eqnarray}
u^{\text{in}}_k(x,\eta)&=\sigma^{\text{in}}(\eta) \exp\left[i (kx-\omega_{_+}\eta)-\frac{i\omega_-}{\rho}\ln[2\cosh(\rho\eta)] \right]\,, \\
u^{\text{out}}_k(x,\eta)&=\sigma^{\text{out}}(\eta) \exp \left[i (kx-\omega_{_+}\eta)-\frac{i\omega_-}{\rho}\ln[2\cosh(\rho\eta)] \right]\,,
\end{eqnarray}
where
\begin{eqnarray}
\sigma^{\text{in}}(\eta)&=\frac{1}{2\sqrt{\pi \omega_{\text{in}}}}F_1\left[1+\frac{i\omega_-}{\rho},\frac{i\omega_-}{\rho};1-\frac{i\omega_{\text{in}}}{\rho};\frac{1}{2}[1-\tanh(\rho\eta)]\right]\,, \\
\sigma^{\text{out}}(\eta)&=\frac{1}{2\sqrt{\pi \omega_{\text{out}}}}F_1\left[1+\frac{i\omega_-}{\rho},\frac{i\omega_-}{\rho};1-\frac{i\omega_{\text{out}}}{\rho};\frac{1}{2}[1-\tanh(\rho\eta)]\right]\,,
\end{eqnarray}
with $F_1$ the hypergeometric function, and
\begin{equation}
\omega_{\text{in}}=\sqrt{k^2+m^2}\,, \qquad
\omega_{\text{out}}=\sqrt{k^2+m^2(1+2\epsilon)}\,, \qquad
\omega_{\pm}=\frac12(\omega_{\text{out}}\pm \omega_{\text{in}})\,.
\end{equation}

The in and out modes are related by the Bogoliubov transformation\cite{L29,Birrell1982}
\begin{equation}\label{modolibov}
u^{\text{in}}_k(x,\eta)=\alpha_k  u^{\text{out}}_k(x,\eta)+\beta_k u^{{\text{out}*}}_{-k}(x,\eta)\,,
\end{equation}
with Bogoliubov coefficients given by
\begin{eqnarray}\label{alphak}
\alpha_k&=\sqrt{\frac{\omega_{\text{out}}}{\omega_{\text{in}}}}\frac{\Gamma([1-(i\omega_{\text{in}}/\rho)])\Gamma(-i\omega_{\text{out}}/\rho)}{\Gamma([1-(i\omega_{+}/\rho)])\Gamma(-i\omega_{+}/\rho)}\,, \\[3mm]
\label{betak}\beta_k&=\sqrt{\frac{\omega_{\text{out}}}{\omega_{\text{in}}}}\frac{\Gamma([1-(i\omega_{\text{in}}/\rho)])\Gamma(i\omega_{\text{out}}/\rho)}{\Gamma([1+(i\omega_{-}/\rho)])\Gamma(i\omega_{-}/\rho)}\,.
\end{eqnarray}
In accordance with Eq.(\ref{modolibov}), the Bogoliubov relationship between the operators of the in and out modes is
\begin{equation}\label{F4}
b_{\text{in},k}=\alpha_k^*b_{\text{out},k}-\beta_k^*b_{\text{out},{-k}}^\dagger,
\end{equation}
where $b_{\text{in},k}$ is the bosonic annihilation operator acting on the state
in the asymptotic past, and $b_{\text{out},k}$ and $b_{\text{out},{-k}}^\dagger$
are the bosonic annihilation and antibosonic creation operators acting
on the states in the asymptotic future. This allows us to write the in-vacuum state as \cite{L28,L29,L30,Birrell1982}
\begin{equation}\label{F5}
|0_k\rangle_{\text{in}}=\sqrt{1-|\frac{\beta_k}{\alpha_k}|^2}\sum_{n=0}^\infty
(\frac{\beta^*_k}{\alpha^*_k})^n
|n_k\rangle_{\text{out}} |n_{-k}\rangle_{\text{out}},
\end{equation}
where $n_k$ represents the $n$ bosons and $n_{-k}$ represents the $n$ antibosons.
This equation clearly shows that the in-vacuum state evolves into a two-mode entangled state in the asymptotic future. Acting the Hermitian conjugate of Eq.(\ref{F4}) on the above in-vacuum state, one obtains \cite{L31}
\begin{equation}\label{F6}
|1_k\rangle_{\text{in}}=\frac{1}{|\alpha|\alpha^*}\sum_{n=0}^\infty
(\frac{\beta^*_k}{\alpha^*_k})^n\sqrt{n+1}
|(n+1)_k\rangle_{\text{out}} |n_{-k}\rangle_{\text{out}}.
\end{equation}
%------------------------------------------------------------------------------------------------------------------------------------------------------------------------------------------------%
\section{Logarithmic negativity and  mutual information in an expanding spacetime\label{GSCDGE}}
%--------------------------------------------------------------------------------
Consider an entangled state between a qubit $p$ (shared by Alice) and a boson of momentum $k$ (shared by Bob) in the asymptotic past
\begin{equation}\label{F8}
\Psi_{p,k}=\chi|\uparrow\rangle|0_k\rangle_{\text{in}}+\sqrt{1-\chi^2}|\downarrow\rangle|1_k\rangle_{\text{in}},
\end{equation}
where $\chi$ is a state parameter which runs from 0 to 1, and $|\uparrow\rangle$ and $|\downarrow\rangle$ denote the basis states of the qubit.
In the simulation experiment, we consider a single qubit (A)  and the boson (B)  in their respective cavities. The two cavities are assumed to be the
same otherwise except that they are separated by a spacelike distance. Then, we close Alice's cavity through the switching function and turn on Bob's switching function.
When the boson evolves from the asymptotic past to  the asymptotic future in a cavity, the bipartite state becomes mixed and can be measured by the decoherence factor.
Using Eqs.(\ref{F5}) and (\ref{F6}), we can rewrite the entangled state as
\begin{equation}\label{F9}
\Psi_{p,k}=\frac{1}{|\alpha_k|}\sum_{n=0}^\infty
(\frac{\beta^*_k}{\alpha^*_k})^n
[\chi|\uparrow\rangle|n_k\rangle_{\text{out}} |n_{-k}\rangle_{\text{out}}+\sqrt{1-\chi^2}\frac{\sqrt{n+1}}{\alpha_k^*}|\downarrow\rangle|(n+1)_k\rangle_{\text{out}} |n_{-k}\rangle_{\text{out}}].
\end{equation}
In this way, the entangled state encodes historical information concerning the expanding spacetime.
We can understand the properties of the underlying spacetime
structure by analysing the quantum properties of this state, such as logarithmic negativity and  mutual information.
For simplicity, we omit the subscript ``out" below.

\subsection{Logarithmic negativity and  mutual information between qubit and boson}\label{subsce3a}
We first study the quantum correlations between the qubit $p$ and the boson with momentum $k$. By tracing over the antiboson mode $-k$, we obtain the density operator $\rho_{p,k}$ for the system of qubit and boson
\begin{eqnarray}\label{F11}
\nonumber\rho_{p,k}&=&\frac{\chi^2}{|\alpha_k|^2}\sum_{n=0}^\infty\gamma^{2n}|\uparrow,n_k\rangle \langle\uparrow,n_k|+ \frac{1-\chi^2}{|\alpha_k|^4}\sum_{n=0}^\infty\gamma^{2n}(n+1)|\downarrow,(n+1)_k\rangle \langle\downarrow,(n+1)_k| \\
\nonumber &+&\frac{\chi\sqrt{1-\chi^2}}{|\alpha_k|^2\alpha_k}\sum_{n=0}^\infty\gamma^{2n}\sqrt{n+1}|\uparrow,n_k\rangle
\langle\downarrow,(n+1)_k| \\
&+&\frac{\chi\sqrt{1-\chi^2}}{|\alpha_k|^2\alpha_k^*}\sum_{n=0}^\infty\gamma^{2n}\sqrt{n+1}
|\downarrow,(n+1)_k\rangle\langle\uparrow,n_k|,
\end{eqnarray}
with $ \gamma^2=|\frac{\beta_k}{\alpha_k}|^2=
\frac{\sinh^2(\pi\frac{\omega_-}{\rho})}
{\sinh^2(\pi\frac{\omega_+}{\rho})}$.

There are many special approaches to the detection of quantum entanglement, among them the Peres-Horodecki positive partial transpose (PPT) criterion \cite{Peres1996,Horodecki1996} and the afterwards logarithmic negativity\cite{L32} are widely used, due to the simplicity of calculations. In this paper, we use the logarithmic negativity as the detection of entanglement, which is defined as
$N(\rho)=\log_{2}\|\rho^{T}\|_{1}$, where the trace norm is given by $\|A\|_{1}={\rm tr}\sqrt{A^{\dag}A}$ and $\rho^{T}$ denotes the partial transpose of the bipartite density matrix $\rho$ with respect to its one party. It is worthwhile to point out that logarithmic negativity is only the sufficient condition of entanglement and is not necessarily a guarantee of detecting all the entanglement in any case. However, it is indeed a feasible indicator for the conjecture of universe expansion.

Performing partial transpose on density matrix $\rho_{p,k}$ with respect to the qubit, then the transposed density matrix has the eigenvalues
\begin{equation}\label{F12}
\frac{\chi^2}{|\alpha_k|^2}, ~~~~\frac{\gamma^{2n}}{2|\alpha_k|^2}\left[\gamma^2\chi^2+\frac{n(1-\chi^2)}{|\beta_k|^2} \pm
\sqrt{\left(\gamma^2\chi^2+\frac{n(1-\chi^2)}{|\beta_k|^2}\right)^2+\frac{4\chi^2(1-\chi^2)}{|\alpha_k|^2}}    \right]\nonumber
\end{equation}
with $ n=0,1,2, \cdots $.
Due to $\omega_-<\omega_+$, thus $\gamma^2<1$ and $|\alpha_k|^2=\frac{1}{1-\gamma^2}\neq\infty$, which means that there must be negative eigenvalues and the state $\rho_{p,k}$ is always entangled.
The logarithmic negativity $N(\rho_{p,k})$ of the state $\rho_{p,k}$ is
\begin{equation}\label{F13}
N(\rho_{p,k})=\log_2\left[\frac{\chi^2}{|\alpha_k|^2}+\sum_{n=0}^{\infty}\frac{\gamma^{2n}}{|\alpha_k|^2}
\sqrt{\left(\gamma^2\chi^2+\frac{n(1-\chi^2)}{|\beta_k|^2}\right)^2+\frac{4\chi^2(1-\chi^2)}{|\alpha_k|^2}}     \right].
\end{equation}

We can also study the total correlation in the state $\rho_{p,k}$ by calculating the mutual
information, which is defined as \cite{L33}
\begin{equation}\label{F15}
I(\rho_{p,k})=S(\rho_{p})+S(\rho_{k})-S(\rho_{p,k}),
\end{equation}
where $S(\rho)=-\rm Tr[\rho\log_2(\rho)]$ is the von Neumann entropy of
the density matrix $\rho$. From the expression $\rho_{p,k}$ of Eq.(\ref{F11}), we obtain the von Neumann entropies for the joint system and each subsystems as
\begin{equation}\label{F115}
S(\rho_{p,k})=-\sum_{n=0}^{\infty}\frac{\gamma^{2n}}{|\alpha_k|^2}\left(\chi^2+ \frac{(n+1)(1-\chi^2)}{|\alpha_k|^2}\right)\log_2\left[\frac{\gamma^{2n}}{|\alpha_k|^2}\left(\chi^2+ \frac{(n+1)(1-\chi^2)}{|\alpha_k|^2}\right)\right],
\end{equation}
\begin{equation}\label{F16}
S(\rho_{p})=-\chi^2\log_2\left[\chi^2\right]-(1-\chi^2)\log_2\left[(1-\chi^2)\right],
\end{equation}
and
\begin{equation}\label{F125}
S(\rho_{k})=-\sum_{n=0}^{\infty}\frac{\gamma^{2n}}{|\alpha_k|^2}\left(\chi^2+ \frac{n(1-\chi^2)}{|\beta_k|^2}\right)\log_2\left[\frac{\gamma^{2n}}{|\alpha_k|^2}\left(\chi^2+ \frac{n(1-\chi^2)}{|\beta_k|^2}\right)\right],
\end{equation}
respectively. Due to $\gamma^2<1$, we thus have $S(\rho_{k})>S(\rho_{p,k})$ and $I(\rho_{p,k})>S(\rho_{p})$.

Eqs.(\ref{F13}) and (\ref{F15}) suggest that logarithmic negativity and mutual information depend on the
cosmological parameters $\epsilon$  and $\rho$, as well as the bosonic mass $m$ and momentum $k$.
Assuming that we live in the universe corresponding to the asymptotic future of the spacetime. Then we can find the  historical information about the spacetime expansion by inspecting the evolution of logarithmic negativity and mutual information.
The historical information shows a promising application in observational cosmology and presents a way to simulate the expansion of universe.

\begin{figure}
\begin{minipage}[t]{0.5\linewidth}
\centering
\includegraphics[width=3.0in,height=5.2cm]{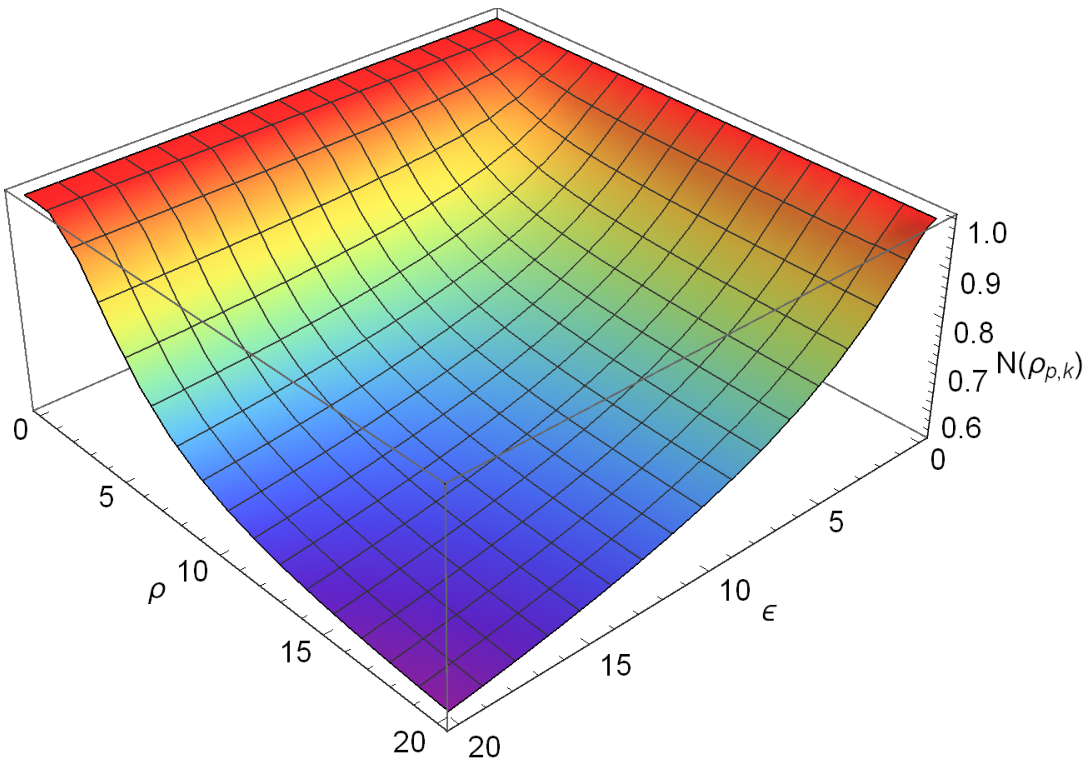}
\label{fig1a}
\end{minipage}%
\begin{minipage}[t]{0.5\linewidth}
\centering
\includegraphics[width=3.0in,height=5.2cm]{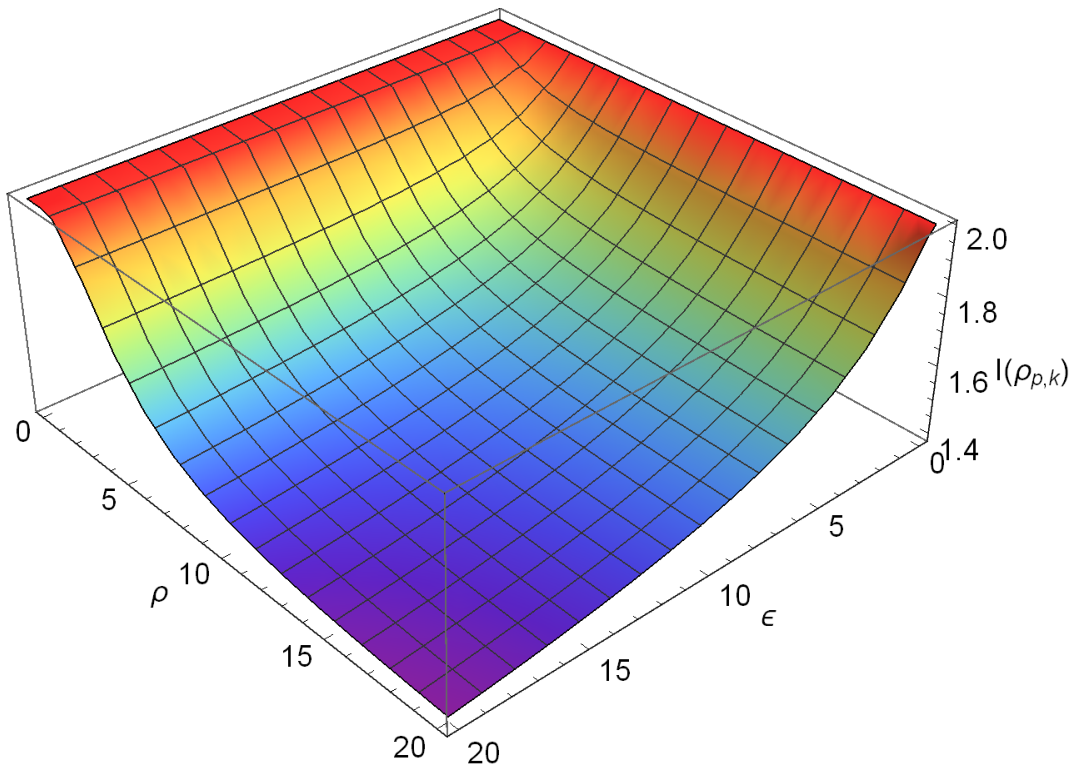}
\label{fig1c}
\end{minipage}%
\caption{ Logarithmic negativity  $N(\rho_{p,k})$ (left) and mutual information $I(\rho_{p,k})$ (right) as functions of the expansion volume $\epsilon$ and the expansion rate $\rho$, where $\chi=\frac{1}{\sqrt{2}}$ and $m=k=1$.}
\label{Fig1}
\end{figure}

In Fig.\ref{Fig1}, we show the logarithmic negativity (left panel) and mutual information (right panel) between qubit and boson as functions of the expansion volume $\epsilon$ or the expansion rate $\rho$. It is shown that the change of mutual information $I(\rho_{p,k})$ is very similar to the change of logarithmic negativity $N(\rho_{p,k})$. They reduce monotonically with respect to both the expansion volume $\epsilon$ and the expansion rate $\rho$.

\begin{figure}
\begin{minipage}[t]{0.5\linewidth}
\centering
\includegraphics[width=3.0in,height=5.2cm]{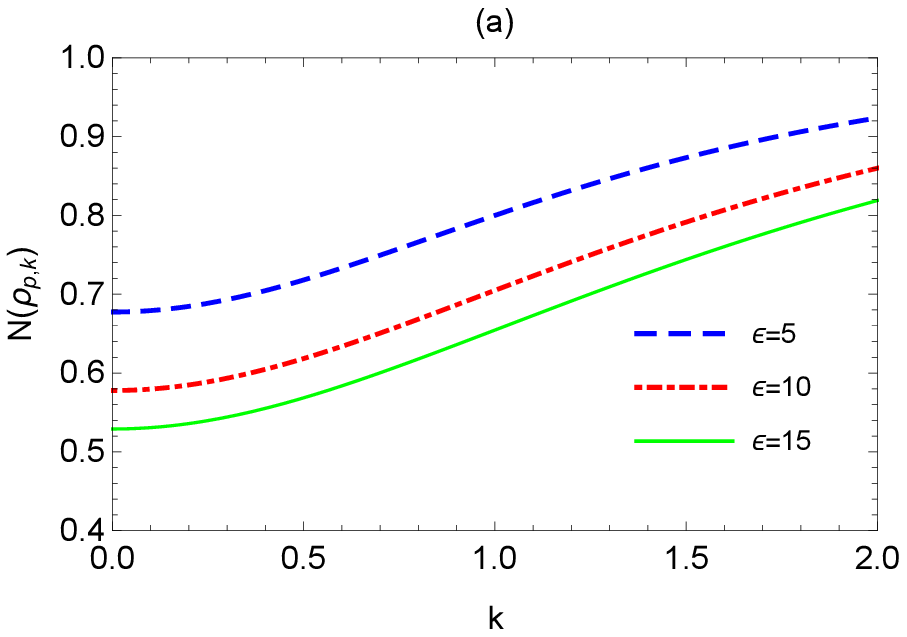}
\end{minipage}%
\begin{minipage}[t]{0.5\linewidth}
\centering
\includegraphics[width=3.0in,height=5.2cm]{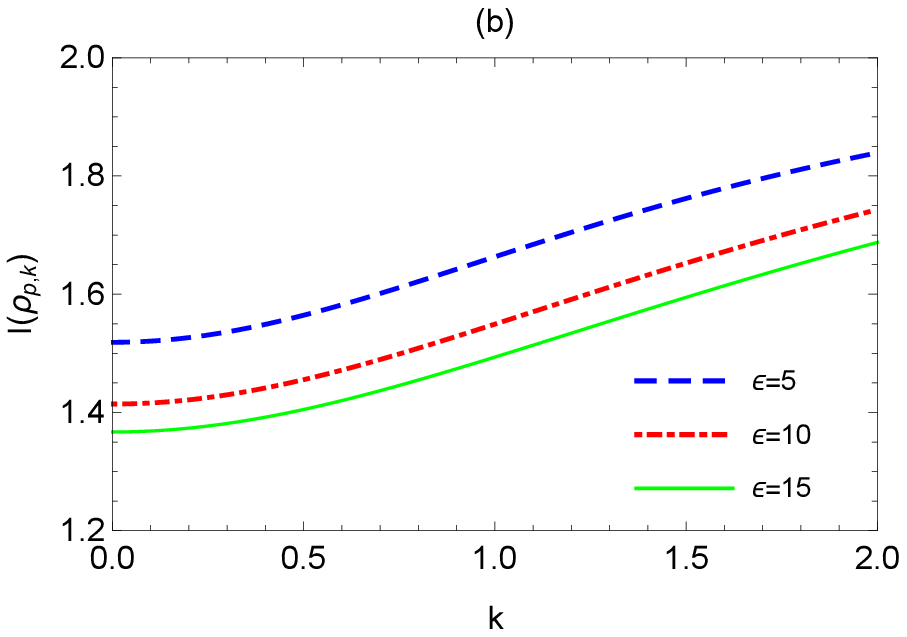}
\end{minipage}%
\vspace{0.5cm}
\begin{minipage}[t]{0.5\linewidth}
\centering
\includegraphics[width=3.0in,height=5.2cm]{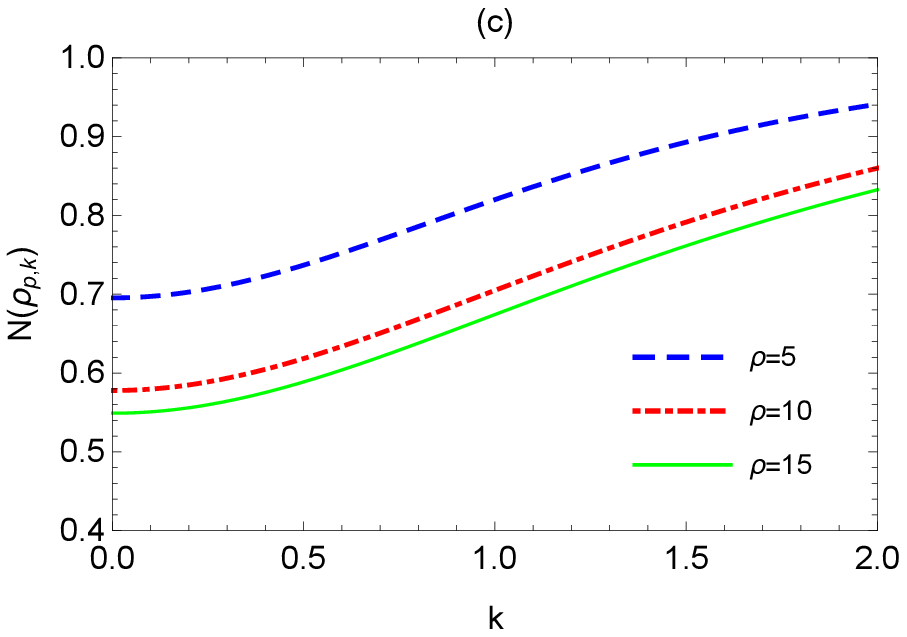}
\end{minipage}%
\begin{minipage}[t]{0.5\linewidth}
\centering
\includegraphics[width=3.0in,height=5.2cm]{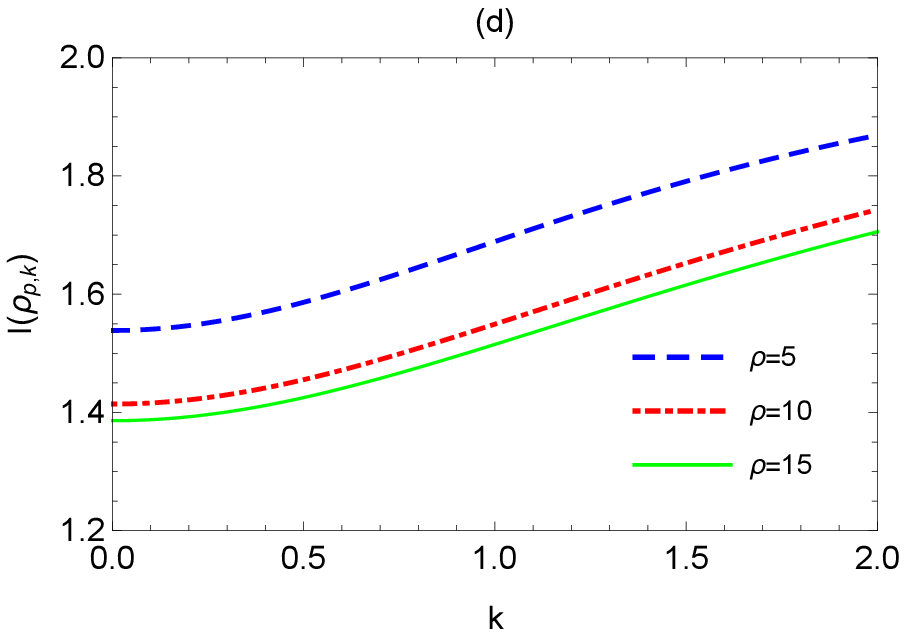}
\end{minipage}%
\caption{Logarithmic negativity  $N(\rho_{p,k})$ (left) and mutual information $I(\rho_{p,k})$ (right) as functions of  the momentum $k$ for different expansion volume $\epsilon$ or different expansion rate $\rho$.  Where we have
fixed $m=1$ and $\chi=\frac{1}{\sqrt{2}}$. $\rho=10$ in (a) and (b), and $\epsilon=10$ in (c) and (d).}
\label{Fig2}
\end{figure}

Fig.\ref{Fig2} shows the dependency of logarithmic negativity and mutual information on the momentum $k$, where the expansion volume $\epsilon$ and expansion rate $\rho$ are set to be three typical values according to Fig.\ref{Fig1}. We find that both the logarithmic negativity $N(\rho_{p,k})$ and the mutual information $I(\rho_{p,k})$ increase monotonically with the growth of momentum $k$. For infinite $k$, we have
$$\lim_{k\rightarrow \infty}N(\rho_{p,k})=\log_2\left[1+2\chi\sqrt{1-\chi^2} \right], ~~~~\lim_{k\rightarrow \infty}I(\rho_{p,k})=2S(\rho_{p}).$$ Note that $\log_2\left[1+2\chi\sqrt{1-\chi^2} \right]$ and $2S(\rho_{p})$ are respectively the initial values of logarithmic negativity and mutual information in the state of Eq.(\ref{F8}), we thus conclude that, for boson with very large momentum, its logarithmic negativity and mutual information with the qubit are very little affected by the expanding spacetime. If we want to extract historical information about the expanding spacetime, the boson modes with smaller momentum are most preferable.

\begin{figure}
\begin{minipage}[t]{0.5\linewidth}
\centering
\includegraphics[width=3.0in,height=5.2cm]{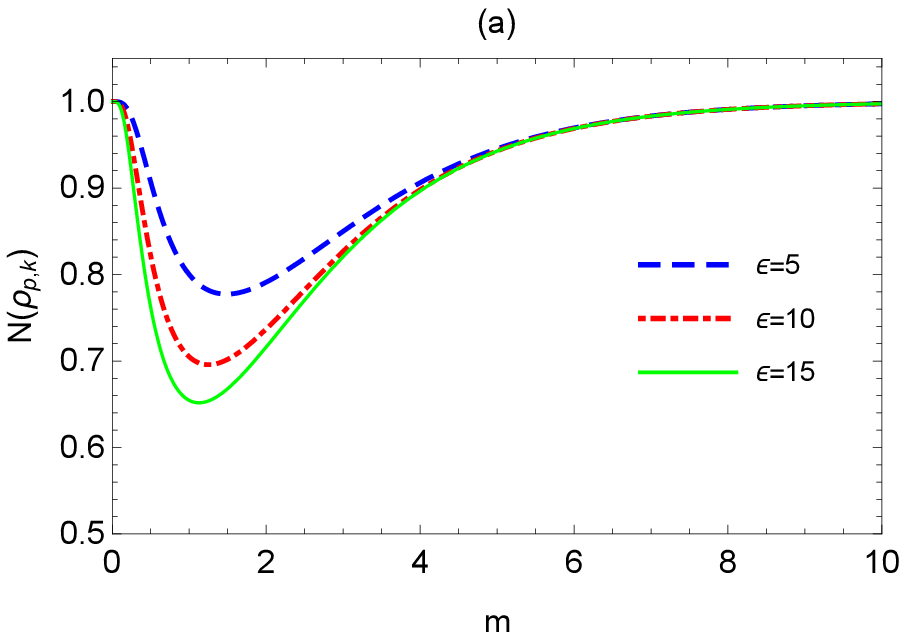}
\end{minipage}%
\begin{minipage}[t]{0.5\linewidth}
\centering
\includegraphics[width=3.0in,height=5.2cm]{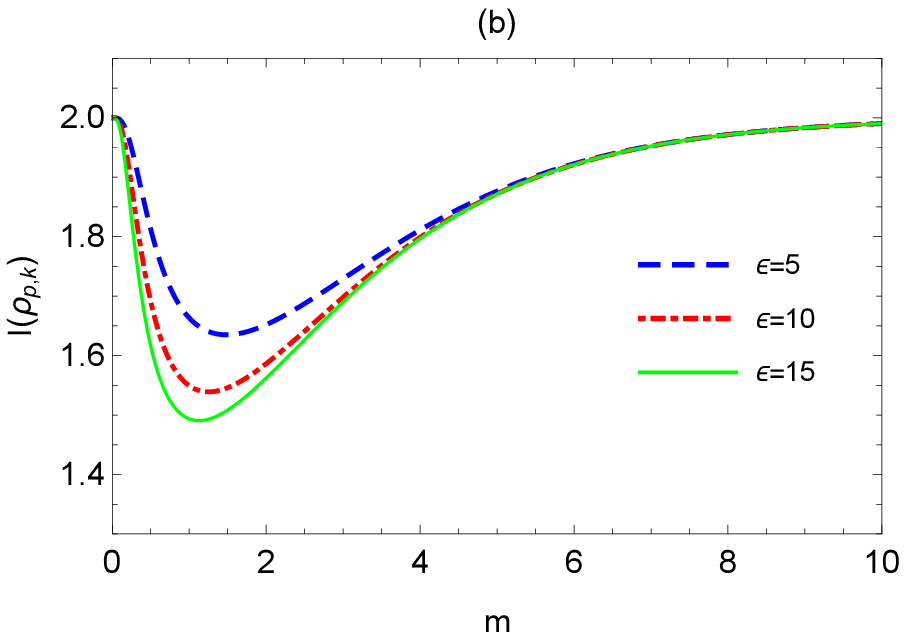}
\end{minipage}%
\vspace{0.5cm}
\begin{minipage}[t]{0.5\linewidth}
\centering
\includegraphics[width=3.0in,height=5.2cm]{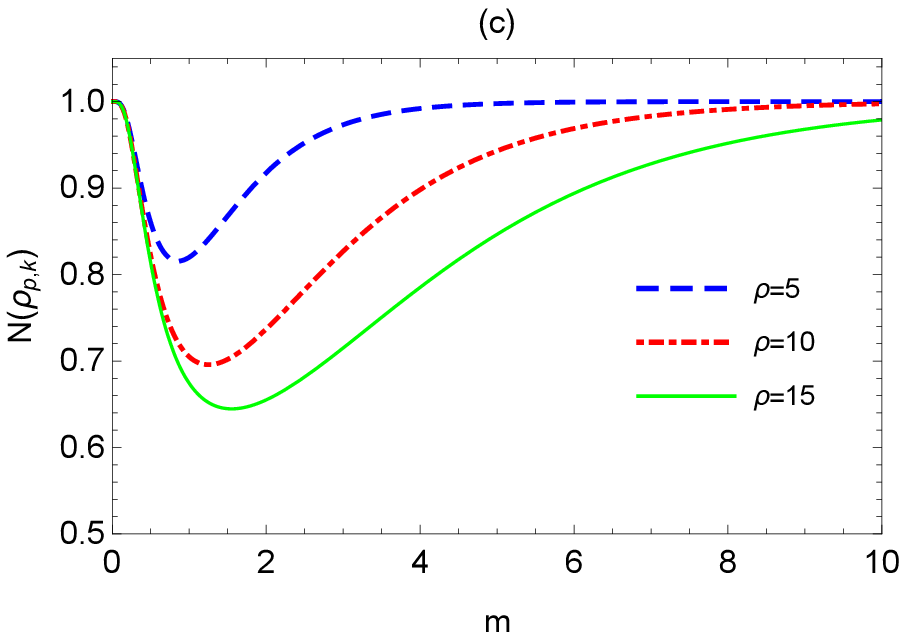}
\end{minipage}%
\begin{minipage}[t]{0.5\linewidth}
\centering
\includegraphics[width=3.0in,height=5.2cm]{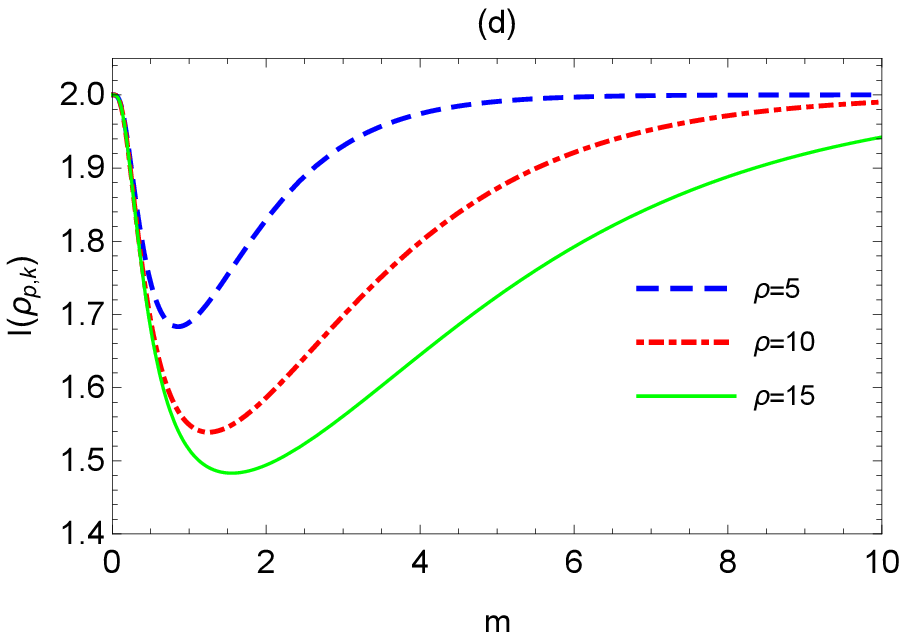}
\end{minipage}%
\caption{Logarithmic negativity $N(\rho_{p,k})$ (left) and mutual information $I(\rho_{p,k})$ (right) as functions of the mass $m$ for different expansion volume $\epsilon$ or different expansion rate $\rho$.  Where we have
fixed $k=1$ and $\chi=\frac{1}{\sqrt{2}}$. $\rho=10$ in (a) and (b), and $\epsilon=10$ in (c) and (d).}
\label{Fig3}
\end{figure}

Fig.\ref{Fig3} shows the change of logarithmic negativity and mutual information versus the mass $m$ for different expansion volume $\epsilon$ and different expansion rate $\rho$.
It is shown that logarithmic negativity $N(\rho_{p,k})$ and mutual information $I(\rho_{p,k})$ firstly decrease to the minimum and then asymptotically recover to their initial values with the increasing of mass $m$. The fact that logarithmic negativity and mutual information asymptotically recover to their initial values can also be demonstrate through the analytical calculations
$$\lim_{m\rightarrow 0}N(\rho_{p,k})=\lim_{m\rightarrow \infty}N(\rho_{p,k})=\log_2\left[1+2\chi\sqrt{1-\chi^2} \right],$$
$$\lim_{m\rightarrow 0}I(\rho_{p,k})=\lim_{m\rightarrow \infty}I(\rho_{p,k})=2S(\rho_{p}).$$
This result suggests that logarithmic negativity and mutual information between the qubit and the boson with very large or very small mass are less affected by the expanding spacetime. There is an optimal mass that is most prone to extract the information about the expanding spacetime. Note that the optimal mass decreases slightly with the growth of the expansion volume $\epsilon$, and increases slightly with the growth of the expansion rate $\rho$.

\subsection{Logarithmic negativity and  mutual information between qubit and antiboson}
In order to study the redistribution of quantum correlations, we need to calculate logarithmic negativity and mutual information between qubit $p$ and the antiboson $-k$. Tracing over the mode $k$ in state of Eq.(\ref{F9}), we obtain the density
matrix
\begin{eqnarray}\label{F18}
\nonumber\rho_{p,-k}&=&\frac{\chi^2}{|\alpha_k|^2}\sum_{n=0}^\infty\gamma^{2n}|\uparrow,n_{-k}\rangle \langle\uparrow,n_{-k}|+ \frac{1-\chi^2}{|\alpha_k|^4}\sum_{n=0}^\infty\gamma^{2n}(n+1)|\downarrow,n_{-k}\rangle \langle\downarrow,n_{-k}| \\ \nonumber
&+&\frac{\chi\sqrt{1-\chi^2}\beta_k^*}{|\alpha_k|^4}\sum_{n=0}^\infty\gamma^{2n}\sqrt{n+1}|\uparrow,(n+1)_{-k}\rangle
\langle\downarrow,n_{-k}| \\
&+&\frac{\chi\sqrt{1-\chi^2}\beta_k}{|\alpha_k|^4}\sum_{n=0}^\infty\gamma^{2n}\sqrt{n+1}
|\downarrow,n_{-k}\rangle\langle\uparrow,(n+1)_{-k}|.
\end{eqnarray}
The partial transpose of this density matrix with respect to qubit has the eigenvalues
\begin{eqnarray}\label{F19}
\frac{1-\chi^2}{|\alpha_k|^4},~~~~ \frac{1}{2|\alpha_k|^2}\gamma^{2n}\left[D \pm \sqrt{D^2-\frac{4\chi^2(1-\chi^2)|\beta_k|^2}{|\alpha_k|^4}}    \right]\nonumber
\end{eqnarray}
with $D=\chi^2+\frac{(n+2)(1-\chi^2)|\beta_k|^2}{|\alpha_k|^4}$ and $n=0,1,2, \cdots. $
Thus the logarithmic negativity between the qubit and antiboson is
\begin{equation}\label{F20}
N(\rho_{p,-k})=\log_2\left[\frac{1-\chi^2}{|\alpha_k|^4}+\sum_{n=0}^{\infty}\frac{\gamma^{2n}}{|\alpha_k|^2}
\left(\chi^2+\frac{(n+2)(1-\chi^2)|\beta_k|^2}{|\alpha_k|^4}\right)     \right]=0.
\end{equation}
This means that spacetime expanding can not generate logarithmic negativity between the qubit and antiboson. However, this does not mean that quantum state between qubit and antiboson
is not without quantum correlation. The correlated quantum coherence  is nonlocal coherence and  is regarded as a kind of quantum correlation. Through simple calculation, we find that the correlated quantum coherence can be generated by the spacetime expanding \cite{wsm32}.

The mutual information between the qubit and the antiboson is given by
\begin{equation}\label{F21}
I(\rho_{p,-k})=S(\rho_{p})+S(\rho_{-k})-S(\rho_{p,-k}),
\end{equation}
where $S(\rho_{p})$ is given by Eq.(\ref{F16}), and $S(\rho_{-k})$ and $S(\rho_{p,-k})$ are given by
\begin{equation}\label{F22}
S(\rho_{-k})=
-\sum_{n=0}^{\infty}\frac{\gamma^{2n}}{|\alpha_k|^2}\left(\chi^2+ \frac{(n+1)(1-\chi^2)}{|\alpha_k|^2}\right)\log_2\left[\frac{\gamma^{2n}}{|\alpha_k|^2}\left(\chi^2+ \frac{(n+1)(1-\chi^2)}{|\alpha_k|^2}\right)\right],\nonumber
\end{equation}
and
\begin{equation}\label{F23}
S(\rho_{p,-k})=-\sum_{n=0}^{\infty}\frac{\gamma^{2n}}{|\alpha_k|^2}\left(\chi^2+ \frac{n(1-\chi^2)}{|\beta_k|^2}\right)\log_2\left[\frac{\gamma^{2n}}{|\alpha_k|^2}\left(\chi^2+ \frac{n(1-\chi^2)}{|\beta_k|^2}\right)\right].\nonumber
\end{equation}
Eq.(\ref{F21}) is positive in general, meaning that spacetime expanding can generate mutual information between the qubit and the antiboson.

\begin{figure}
\begin{minipage}[t]{0.5\linewidth}
\centering
\includegraphics[width=3.0in,height=5.2cm]{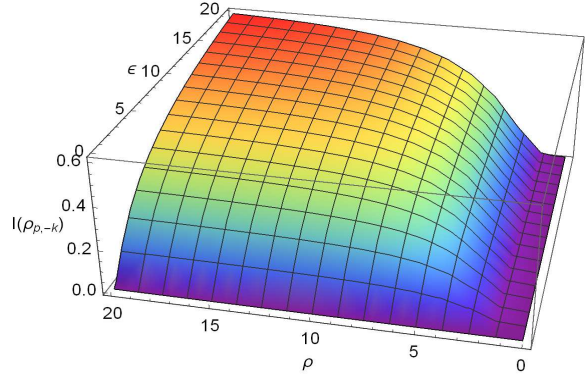}
\end{minipage}%
\caption{Mutual information $I(\rho_{p,-k})$ as a function of the expansion volume $\epsilon$  and the expansion rate $\rho$, where $m=k=1$ and $\chi=\frac{1}{\sqrt{2}}$.}
\label{Fig4}
\end{figure}

In Fig.\ref{Fig4}, we plot the mutual information between the qubit and the antiboson as a function of the expansion volume $\epsilon$ and the expansion rate $\rho$. It is shown that the mutual information $I(\rho_{p,-k})$ increases monotonically with both $\epsilon$ and $\rho$, but the change is more sensitive to the beginning of $\epsilon$ and $\rho$. This means that smaller $\epsilon$ and $\rho$ are more favorable from the consideration of extracting information.

\begin{figure}
\begin{minipage}[t]{0.5\linewidth}
\centering
\includegraphics[width=3.0in,height=5.2cm]{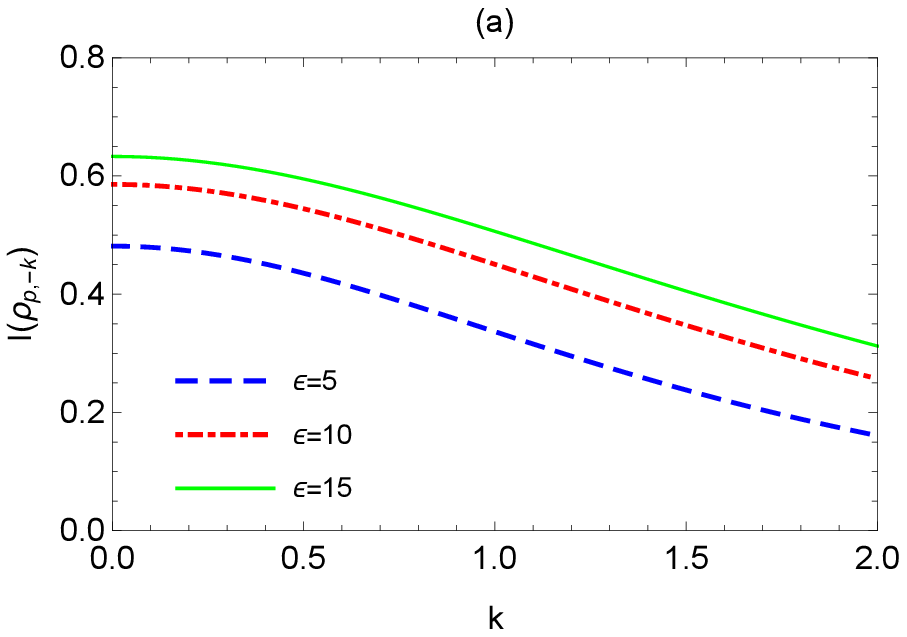}
\end{minipage}%
\begin{minipage}[t]{0.5\linewidth}
\centering
\includegraphics[width=3.0in,height=5.2cm]{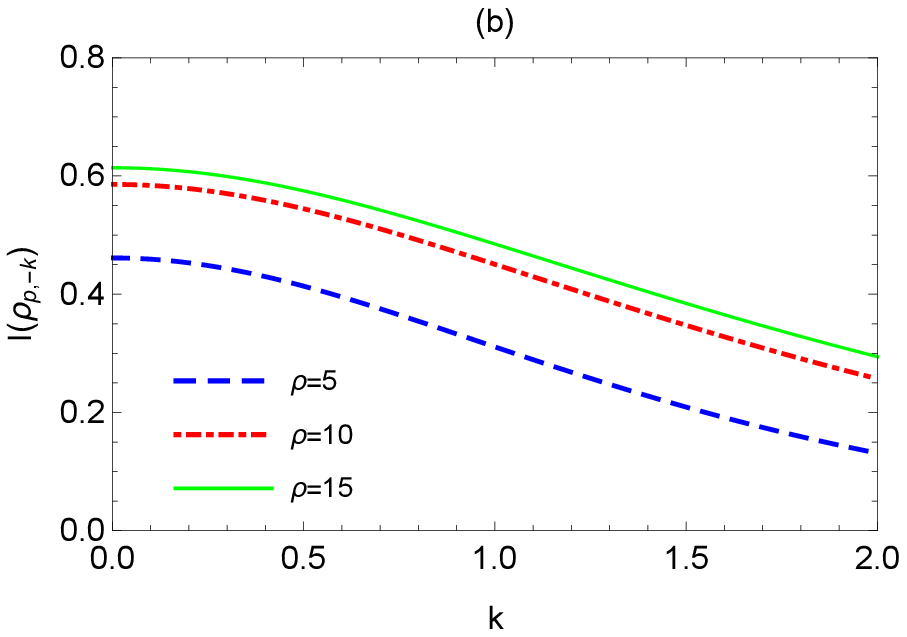}
\end{minipage}%
\vspace{0.5cm}
\begin{minipage}[t]{0.5\linewidth}
\centering
\includegraphics[width=3.0in,height=5.2cm]{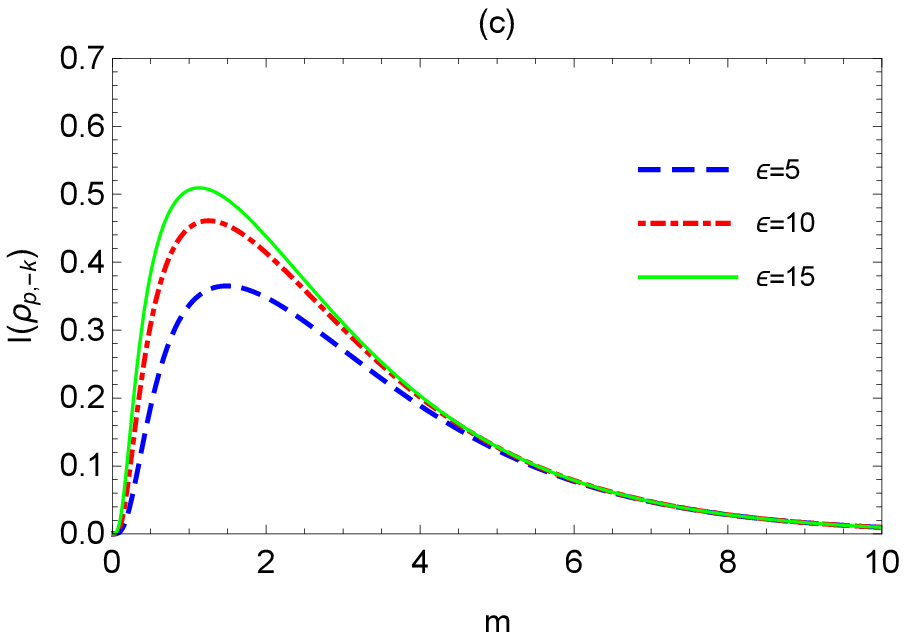}
\end{minipage}%
\begin{minipage}[t]{0.5\linewidth}
\centering
\includegraphics[width=3.0in,height=5.2cm]{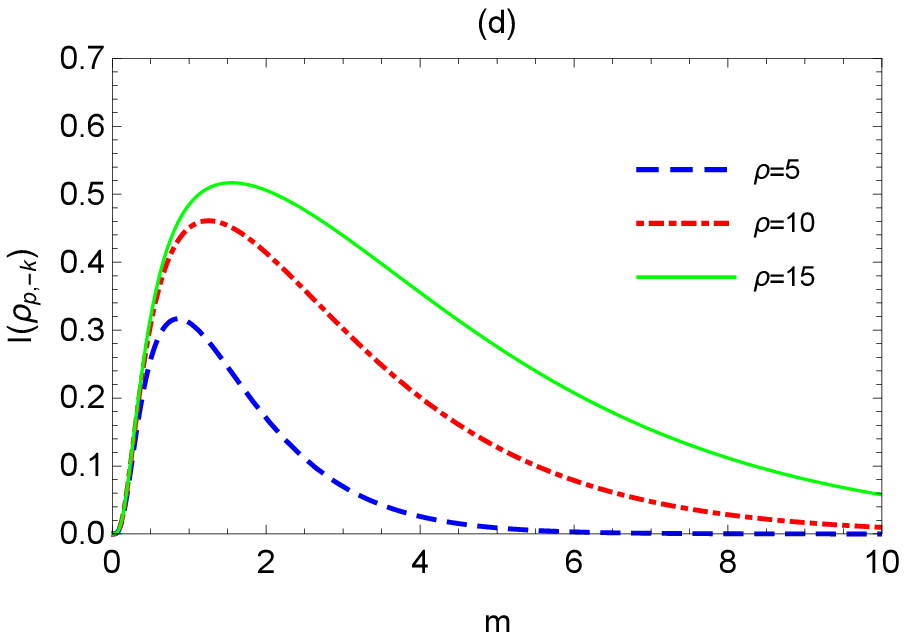}
\end{minipage}%
\caption{Mutual information $I(\rho_{p,-k})$ as functions of  the momentum $k$ or the mass $m$ for different expansion volume $\epsilon$ or different expansion rate $\rho$, where $\chi=\frac{1}{\sqrt{2}}$. Other parameters are chosen as: (a) $m=1$ and $\rho=10$; (b) $m=1$ and $\epsilon=10$; (c) $k=1$ and $\rho=10$; (d) $k=1$ and $\epsilon=10$.}
\label{Fig5}
\end{figure}

Fig.\ref{Fig5} shows the change of mutual information $I(\rho_{p,-k})$  versus the momentum $k$ or mass $m$, where the expansion volume $\epsilon$ and expansion rate $\rho$ are set as the typical values according to Fig.\ref{Fig4}.  We can see that $I(\rho_{p,-k})$ reduces monotonically with the growth of the momentum $k$ [see Fig.\ref{Fig5}(a) and \ref{Fig5}(b)].
In the limit of infinite momentum, we have
$ \lim_{k\rightarrow \infty}I(\rho_{p,-k})=0$ (through analytic calculation).  This means that the
smaller the momentum we pick out, the more the generated mutual information $I(\rho_{p,-k})$ is, which is more favourable for extracting information of spacetime expansion.

From Fig.\ref{Fig5}(c) and \ref{Fig5}(d), we see that the generated mutual information $I(\rho_{p,-k})$ increases from zero to a maximum and then reduces again to zero with the growth of the mass $m$, i.e., satisfying
$$\lim_{m\rightarrow 0}I(\rho_{p,-k})=\lim_{m\rightarrow \infty}I(\rho_{p,-k})=0.$$
This means that there are some medium-sized masses which can produce more mutual information between the qubit and the antiboson and thus are more conducive to extract information about the expanding of universe. As displayed by Fig.\ref{Fig3}, the optimal mass decreases with the growth of the expansion volume $\epsilon$, and increases with the growth of the expansion rate $\rho$. In fact, we can demonstrate (see \ref{sec3c} below) that the optimal mass displayed here is actually consistent with that in Fig.\ref{Fig3}.

\begin{figure}
\begin{minipage}[t]{0.5\linewidth}
\centering
\includegraphics[width=3.0in,height=5.2cm]{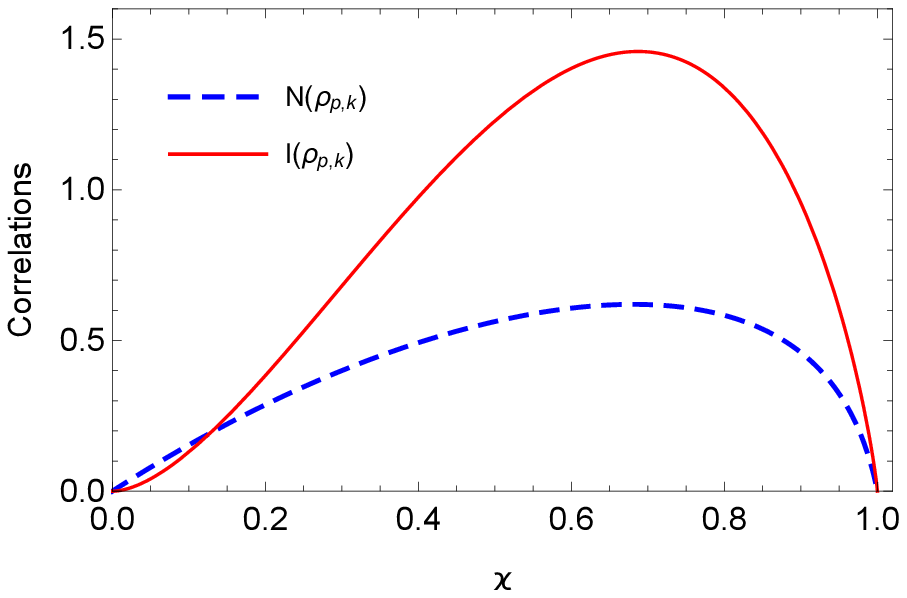}
\label{fig1a}
\end{minipage}%
\begin{minipage}[t]{0.5\linewidth}
\centering
\includegraphics[width=3.0in,height=5.2cm]{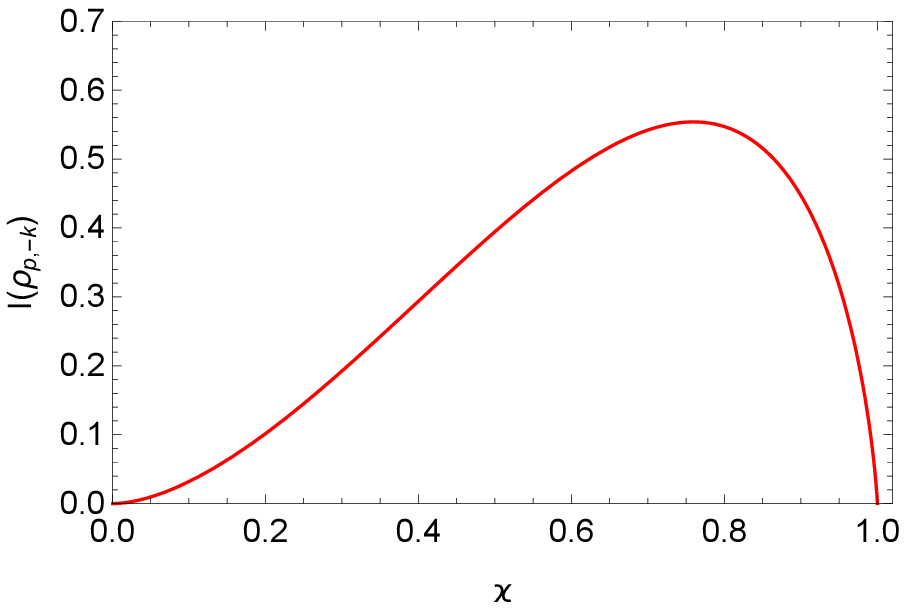}
\label{fig1c}
\end{minipage}%
\caption{Logarithmic negativity and mutual information as functions of the initial parameter $\chi$, where $\epsilon=\rho=10$, $k=0.5$ and $m=1$.}
\label{Fig6}
\end{figure}

\subsection{Trigger and monogamy of mutual information}\label{sec3c}
From the above discussion, we find that the expansion of universe can not produce logarithmic negativity $N(\rho_{p,-k})$, but can produce mutual information $I(\rho_{p,-k})$ between the qubit and the antiboson. However, the production of $I(\rho_{p,-k})$ requires some initial conditions.
In Fig.\ref{Fig6}, we plot the logarithmic negativity $N(\rho_{p,k})$, the mutual information $I(\rho_{p,k})$ and $I(\rho_{p,-k})$ as functions of the initial parameter $\chi$. It is shown that the production of  the mutual information $I(\rho_{p,-k})$ needs the trigger of the initial logarithmic negativity $N(\rho_{p,k})$ or initial mutual information $I(\rho_{p,k})$. With no these triggers, the mutual information $I(\rho_{p,-k})$ can not be generated.
The more the trigger logarithmic negativity $N(\rho_{p,k})$ or the mutual information $I(\rho_{p,k})$ is, the more the generated mutual information $I(\rho_{p,-k})$ is.

In fact, the trigger phenomenon of mutual information is obviously. We can view the boson and antiboson as a bipartite system. If there is no initial mutual information between the qubit and the boson, then also no initial mutual information between the qubit and the bipartite system. As no interaction exists between the qubit and the bipartite system in the expanding of spacetime, the mutual information between them always remains zero. Thus no mutual information exists between the qubit and the antiboson. What's interesting is that the more the trigger mutual information, the more the generated mutual information $I(\rho_{p,-k})$, and this trigger phenomenon does not take place in the logarithmic negativity.

From Figs.\ref{Fig1}-\ref{Fig5}, we find that the changes of $I(\rho_{p,k})$ and $I(\rho_{p,-k})$ are always opposite: Whenever the mutual information between the qubit and the boson $I(\rho_{p,k})$ degrades, then the mutual information between the qubit and the antiboson $I(\rho_{p,-k})$ increases.
After careful inspection of  Eqs.(\ref{F15})
and (\ref{F21}), we find the monogamy relationship
\begin{equation}\label{F30}
I(\rho_{p,k})+I(\rho_{p,-k})=2S(\rho_{p}),
\end{equation}
where $2S(\rho_{p})$ is the initial mutual information for the state of Eq.(\ref{F8}).
This implies that the mutual information is re-distributable: The initial mutual information is redistributed into $\rho_{p,k}$ and $\rho_{p,-k}$, but the total mutual information is conserved in the expansion of universe. This is evidenced by the numerical simulations, i.e., the value of $I(\rho_{p,k})$ in Fig.\ref{Fig3}(b)(d) plus the value of $I(\rho_{p,-k})$ in Fig.\ref{Fig5}(c)(d), for the same $\epsilon$ and $\rho$, is equal to 2. This conservation also demonstrates that the optimal mass appeared in Fig.\ref{Fig3}(b)(d) and Fig.\ref{Fig5}(c)(d) coincide.

\subsection{Entanglement entropy of boson}\label{sec3d}
Ball $et$ $al$ firstly studied the entanglement entropy of bosonic field in an expanding spacetime  \cite{L28}. Then, the comparison between fermionic and bosonic entanglement entropy for free field modes was studied \cite{L8,L29}. It is shown that the more information about cosmological parameters can be extracted by using fermionic  entanglement entropy. Next, Li $et$ $al$ studied decoherence and disentanglement of scalar fields for Unruh-Wald qubit detector model in an expanded spacetime \cite{L31}. The results show that the expansion of the spacetime in its history destroys the
entanglement between the qubits. Based on these work, Moradi $et$ $al$ investigated  spin-particles entanglement between two modes of Dirac field in Robertson-Walker spacetime \cite{wsm33}. The analysis highlight the role of polarization of particles and compare
the polarization results with those obtainable without accounting for it.
Mohammadzadeh $et$ $al$ studied entropy production due to Lorentz invariance violation (LIV) in an expanding spacetime \cite{wsm34}, and Liu $et$ $al$ used the optimal estimation of parameters for scalar fields to extract information in expanding universe exhibiting LIV \cite{L30}. The extracted information about the past existence of LIV might be useful in recovering the quantum structure of gravity. Really, the occurrence of a peak in the behavior of  entanglement entropy for a specific momentum can provide information about the expansion parameters. Therefore, the information about the LIV parameter is codified in this peak.

\begin{figure}
\begin{minipage}[t]{0.5\linewidth}
\centering
\includegraphics[width=3.0in,height=5.2cm]{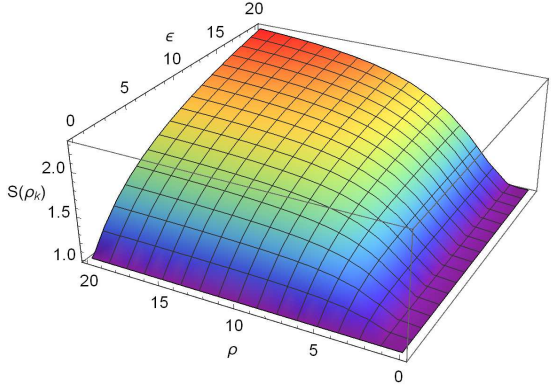}
\end{minipage}%
\caption{Entanglement entropy $S(\rho_{k})$ as a function of the expansion volume $\epsilon$  and the expansion rate $\rho$, where $m=k=1$ and $\chi=\frac{1}{\sqrt{2}}$.}
\label{Fig7}
\end{figure}

In \ref{subsce3a}, we use the logarithmic negativity and mutual information between the qubit and the boson to detect the law of universe expansion, because they contain the information about universe expansion. In fact, we find from Eq.(\ref{F125}) that the entanglement entropy (i.e. von Neumann entropy) $S(\rho_{k})$ of the boson also contains the information about universe expansion, thus it also can be used to detect the law of universe expansion. As the whole system (qubit, boson and antiboson) is always a pure state in the universe expansion, $S(\rho_{k})$ describes actually the quantum correlation of the boson with the qubit and the antiboson. In Fig.\ref{Fig7}, we plot the entanglement entropy $S(\rho_{k})$ as a function of the expansion volume $\epsilon$  and the expansion rate $\rho$, where the parameters are chosen as the same as in Fig.\ref{Fig4}. It is shown that the evolution of $S(\rho_{k})$ is also similar to that of $I(\rho_{p,-k})$ in Fig.\ref{Fig4}. $S(\rho_{k})$ increases monotonically with both $\epsilon$ and $\rho$, and the change is more sensitive to the beginning of $\epsilon$ and $\rho$, meaning that smaller $\epsilon$ and $\rho$ are more favorable for the extraction of information.

\begin{figure}
\begin{minipage}[t]{0.5\linewidth}
\centering
\includegraphics[width=3.0in,height=5.2cm]{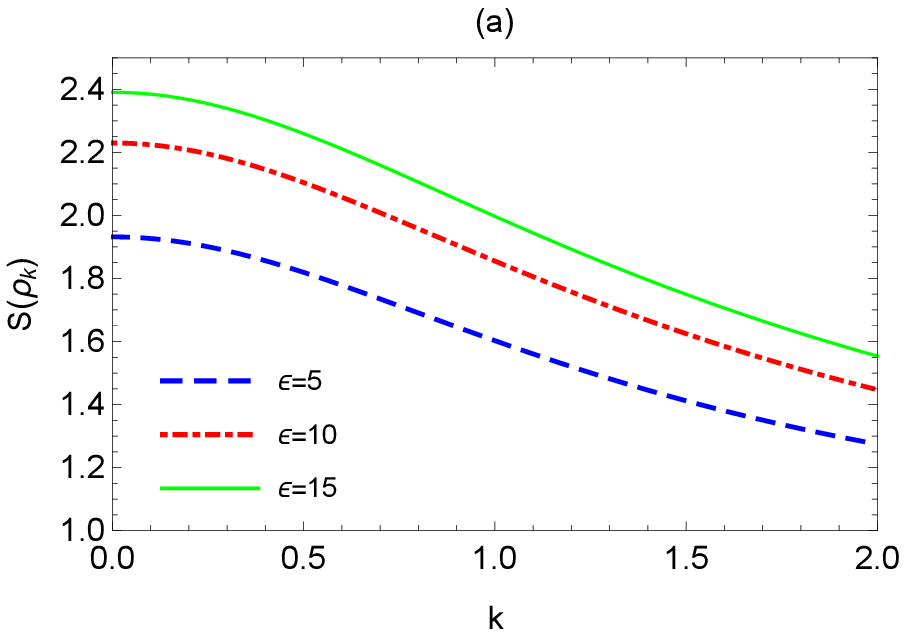}
\end{minipage}%
\begin{minipage}[t]{0.5\linewidth}
\centering
\includegraphics[width=3.0in,height=5.2cm]{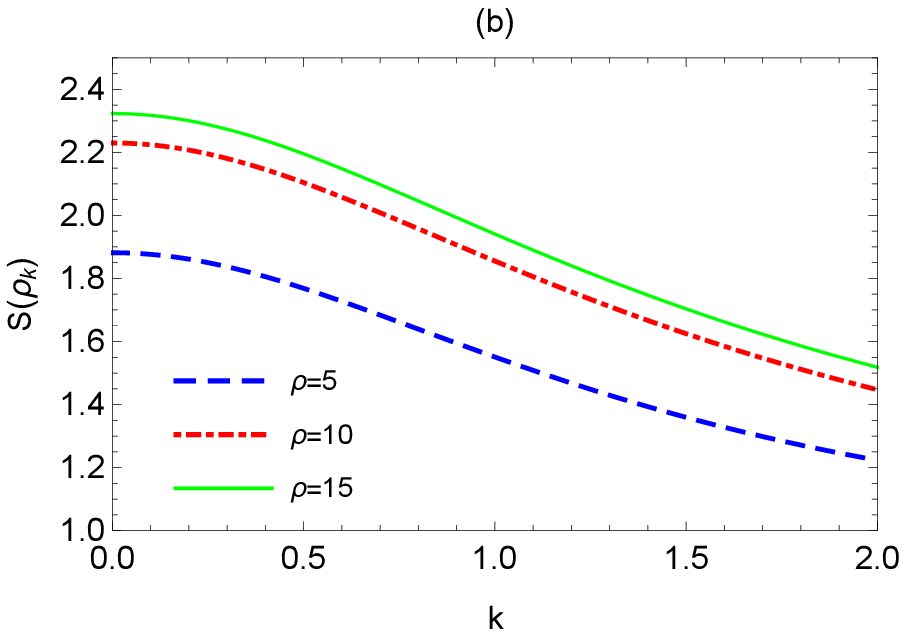}
\end{minipage}%
\vspace{0.5cm}
\begin{minipage}[t]{0.5\linewidth}
\centering
\includegraphics[width=3.0in,height=5.2cm]{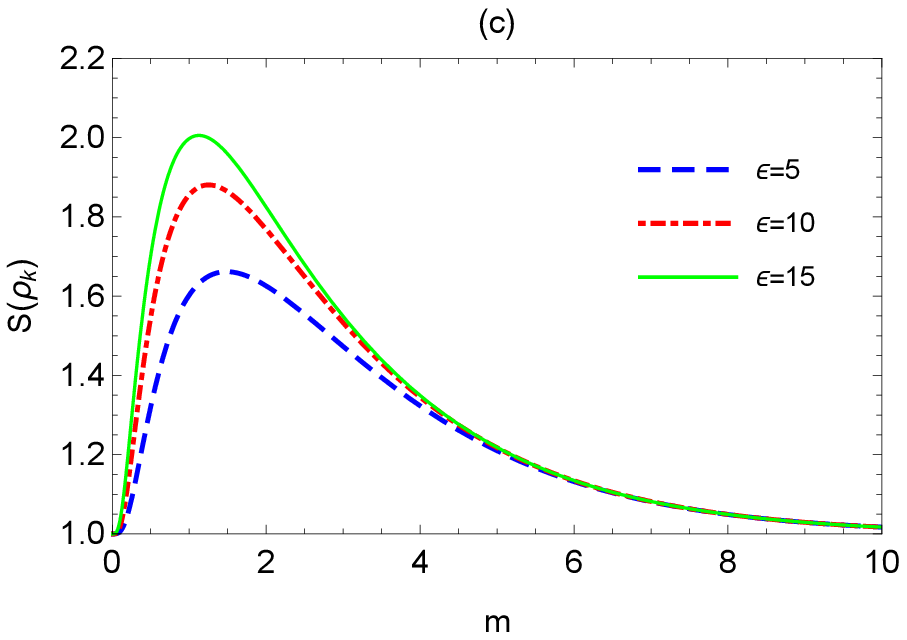}
\end{minipage}%
\begin{minipage}[t]{0.5\linewidth}
\centering
\includegraphics[width=3.0in,height=5.2cm]{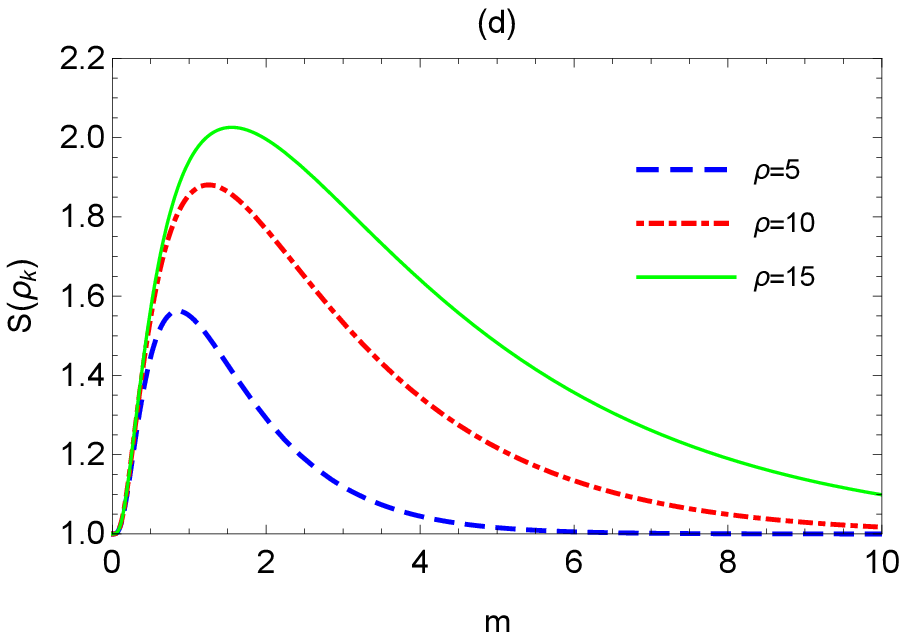}
\end{minipage}%
\caption{Entanglement entropy $S(\rho_{k})$ as functions of  the momentum $k$ or the mass $m$ for different expansion volume $\epsilon$ or different expansion rate $\rho$, where $\chi=\frac{1}{\sqrt{2}}$. Other parameters are chosen as: (a) $m=1$ and $\rho=10$; (b) $m=1$ and $\epsilon=10$; (c) $k=1$ and $\rho=10$; (d) $k=1$ and $\epsilon=10$.}
\label{Fig8}
\end{figure}

Fig.\ref{Fig8} shows the change of entanglement entropy $S(\rho_{k})$ versus the momentum $k$ or mass $m$, where the parameters are chosen as the same as in Fig.\ref{Fig5}. We see that Fig.\ref{Fig8} is also very similar to Fig.\ref{Fig5}. $S(\rho_{k})$ reduces monotonically with the growth of the momentum $k$ [Fig.\ref{Fig8}(a) and \ref{Fig8}(b)]. In the limit of infinite momentum, we have
$ \lim_{k\rightarrow \infty}S(\rho_{k})=0$ (through analytic calculation).  This means that the
smaller the momentum we pick out, the more the generated entanglement entropy $S(\rho_{k})$ is, which is more favourable for extracting information of spacetime expansion.

From Fig.\ref{Fig8}(c) and \ref{Fig8}(d), we see that the generated entanglement entropy $S(\rho_{k})$ increases from zero to a maximum and then reduces again to zero with the growth of the mass $m$, i.e., satisfying
$$\lim_{m\rightarrow 0}S(\rho_{k})=\lim_{m\rightarrow \infty}S(\rho_{k})=0.$$
This means that there are some medium-sized masses which can produce more entanglement entropy for the boson and thus are more conducive to extract information about the universe expansion. Also, the optimal mass decreases slightly with the growth of the expansion volume $\epsilon$, and increases slightly with the growth of the expansion rate $\rho$. However, this optimal mass is conceptually not the same thing as that in Fig.\ref{Fig5}(c)(d) for $I(\rho_{p,-k})$.

The ability to recover information about the cosmological parameters which are given access the generated mutual information between the qubit and the antiboson as well as entanglement entropy, were analysed in the limit of small mass $m\ll2\rho\sqrt{\epsilon}$. In this limit, we find  an approximate relation between the generated mutual information and the mass $m$
$$\gamma^2=|\frac{\beta_k}{\alpha_k}|^2\approx\frac{\epsilon m^2}{2(m^2+k^2)}.$$
We can see that the generated mutual information is a  monotonically increasing
function of the expansion volume $\epsilon$, and is a monotonically decreasing function of the momentum $k$. We can also see that  the generated mutual information is a increasing
function of the mass $m$ for smaller $m$. However, for bigger  mass $m$,  the generated mutual information is a decreasing function of the mass $m$.
In this  same limit, the expansion rate $\rho$ can be related to the variation of the generated mutual information
 $$\rho\approx \pi E\sqrt{\frac{1+\gamma^2}{-E\partial_E\ln[\gamma^2 ]-4}},$$
where $E=\sqrt{m^2+k^2}$ is the energy of the $k$-momentum mode.
Unfortunately, for ${m\rightarrow0}$,  the generated mutual information
is almost zero due to the conformal symmetry of the theory. Therefore, the generated mutual information of massive states in our simple toy
universe encodes the complete information about the cosmological parameters.

In the hopes of stimulating further research into the relation between the generated mutual information and universe expansion, we conclude with some discussion of further simulation avenues in this area.
We discuss the possibility of the cosmic neutrino background that possesses quantum entanglement and mutual information  from
the early universe and surviving all the way to the present because of its weak coupling to
matter. We can also discuss the importance of analogue experimental models providing a
laboratory-accessible testbed for cosmological models. Finally, we can use quantum entanglement and mutual information as a tool for theoretical
and experimental cosmology.

Perhaps, it may be applied in the modeling of universe expansion with superconducting circuit \cite{Zehua2017}.
We theoretically propose a setup to realize Jackiw-Teitelboim
expansion universe as solitons of sine-Gordon. Note that our configuration
can be similar to the superconducting coplanar waveguide. However, each capacitor in
our configuration  is parallel with an identical superconducting
quantum interference devices. It is the added superconducting
quantum interference device providing a nonlinear
potential to the Lagrangian of fluxes in the superconducting coplanar waveguide, which plays
a very crucial role in the simulation of expansion universe.  Each superconducting
quantum interference device  consists of two identical parallel
Josephson junctions. Besides, the geometric size of superconducting
quantum interference devices loop is assumed to be small enough such that its selfinductance
is negligible compared to its kinetic inductance. In this case, each superconducting
quantum interference device can be referred to as an effective
Josephson junction with a junction capacitance and a tunable Josephson energy.
We can realize the Jackiw-Teitelboim
expansion universe as solitons of sine-Gordon equation in our setup. Thus, it
allows us to understand the mechanics of solitons and especially
the duality between Jackiw-Teitelboim gravity and sine-Gordon soliton from the
perspective of experiment. The analogous dynamic behavior of quantum fields under the expanding universe in principle could be implemented in the setup. By appropriately adjusting the external magnetic flux  through the superconducting
quantum interference device loop, we can simulate the behavior of the conformal scale factor. Therefore, the analogous dynamic behavior of quantum fields under an expanding universe in principle could be implemented in setup. The in-mode and out-mode are two particular solutions in the past and future, respectively.
Physically, these two particular solutions are a
description of two different vacuum and excited states in
the two asymptotic regions. We consider the boson (B) that is affected by
the simulated spacetime expanding and the qubit (A) entangled with boson in shielding  environment.

%------------------------------------------------------------------------------------------------------------------------------------------------------------------------------------------------%
\section{ Conclusions  \label{GSCDGE}}
%--------------------------------------------------------------------------------
The effects of the expansion universe on quantum correlations (including logarithmic negativity and mutual information)
between a qubit and a relativistic boson have been investigated.
We have found that the dynamics for both the logarithmic negativity and mutual information between the qubit and the boson are very similar. Both of them reduce with the growth of the expansion rate and the expansion volumes. The degradation of quantum correlations also depends on the nature of the boson. The less the boson momentum is, the more serious the quantum correlations reduce. The dependency of quantum correlations on the mass of boson is however non-monotonic. Very small and very large mass of bosons have little influence, and some medium mass of bosons induce large reduction of quantum correlations.

We have found that, in the expanding of universe, logarithmic negativity between the qubit and the antiboson always remains zero, but the mutual information between them can be generated. With the growth of the expansion rate and the expansion volume, the generated mutual information becomes large and large.
Smaller momentum and medium mass of bosons are more beneficial to generate mutual information between the qubit and the antiboson. Interestingly, the production of the mutual information between the qubit and the antiboson requires a trigger of initial quantum correlations. The more the trigger quantum correlations, the more the generated mutual information.

Further study has suggested that the mutual information is re-distributable. In the expansion of universe, the mutual information between the qubit and the boson reduce, at the same time the mutual information between the qubit and the antiboson is generated, where the total mutual information is conserved. This result is different from logarithmic negativity, which is not re-distributable and has no conservation property.
We have also studied the effect of universe expansion on the entanglement entropy of the boson. With the growth of the expansion rate and the expansion volume, the entanglement entropy of boson is generated and becomes large and large. Smaller momentum and medium mass of bosons are more beneficial for the generation of the entanglement entropy.

In the expansion of universe, quantum correlation dynamics encode the information about the underlying spacetime structure, which suggests a promising application in observational cosmology. The results in this paper suggest that small momentum and medium mass of bosons can more sensitively highlight the changes of quantum correlations and thus are more beneficial for extracting the historical information about the underlying spacetime expansion. Of course, the scheme may be restricted by the current technology, because of the strict assumption that the qubit is shielded from any external influence in the expanding of universe, except for the initial entanglement with the boson. We expect our research can present helps for the relativistic quantum information processing and the observational cosmology.

\begin{acknowledgments}
This work is supported by the National Natural
Science Foundation of China (Grant Nos. 1217050862, 11275064), the Construct Program of the National Key Discipline, and BNUXKJC2017.	
\end{acknowledgments}

$\mathbf{Conflict}$  $\mathbf{of}$ $\mathbf{Interest}$

The authors declare no conflict of interest.

%\newpage

\end{document}